\documentclass[10pt,a4paper,twoside]{article}
\usepackage{epsfig}
\usepackage{baltlat6}
\usepackage{array}
\usepackage{here}
\begin{document}
 \ \ \vspace{0.5mm} \setcounter{page}{000}

 \titlehead{Baltic Astronomy, vol.\,00, 000--000, 2016}
 \titleb
  {ESTIMATION OF THE VERTICAL DISK SCALE HEIGHT USING YOUNG GALACTIC OBJECTS}

\begin{authorl}
 \authorb{V.V. Bobylev}{} and
 \authorb{A.T. Bajkova}{}
\end{authorl}

\begin{addressl}
 \addressb{}{Central (Pulkovo) Astronomical Observatory of RAS, 65/1 Pulkovskoye Ch.,
 St. Petersburg, Russia; bob-v-vzz@rambler.ru}
\end{addressl}


\begin{summary}
We have collected literature data on young Galactic objects such
as masers with VLBI-measured trigonometric parallaxes, OB
associations, HII regions and Cepheids. We have recently
established that vertical disk scale height is strongly influenced
by the objects of the Local arm. In the present work we used
samples that do not contain objects in this arm. Based on the
model of a self-gravitating isothermal disk for the density
distribution, we have found the following vertical disk scale
heights:
 $h=46\pm5$~pc from 69 masers with trigonometric parallaxes,
 $h=36\pm3$~pc from 59 OB associations,
 $h=35.6\pm2.7$~pc from 147 HII regions,
 $h=52.1\pm1.9$~pc from 195 young Cepheids, and
 $h=72.0\pm2.3$~pc from 192 old Cepheids.
  \end{summary}


 \resthead
 {Estimation of the vertical disk scale height using young Galactic objects}
 {V.V. Bobylev, A.T. Bajkova}

\sectionb{1}{INTRODUCTION}
The Galactic thin disk attracts the attention of many authors.
Data on young O- and B-type stars, open clusters, Cepheids,
infrared sources, molecular clouds, and other young objects are
used to study its properties. In particular, such parameters as
the offset of the symmetry plane relative to the Sun $z_\odot$ and
the vertical disk scale height $h$ are important.

Previously (Bobylev \& Bajkova 2016a), we showed the $z$
distribution of Galactic masers with measured trigonometric
parallaxes to be asymmetric. We suggested that this is because the
sample is dominated by the masers observed mainly from the Earth's
northern hemisphere. However, there can also be other sources of
this peculiarity, which should be studied using a large
statistical material. In particular, it is interesting to check
the influence of the Local arm. As we showed previously (Bobylev
\& Bajkova 2014), the symmetry plane of the Local arm has an
inclination of about $6^\circ$ to the Galactic plane. Since about
40 of the 130 masers belong to the Local arm, the effect here can
be significant.

In paper of Bobylev \& Bajkova (2016b) the approach, based on
exclusion of the objects of the Local arm, was applied for
analysis of samples of Galactic objects for which the distances
were determined using the kinematic method. In the present work we
are  going to apply this approach for the analysis of samples, the
distances to which are determined with higher precision, from 10\%
(trigonometric parallaxes) to 15\%--25\% (``period-luminosity'' or
spectrophotometric distances).

\sectionb{2}{METHOD}\label{method}
We use heliocentric rectangular coordinate systems $xyz$ where the
$x$ axis is directed toward the Galactic center, the $y$ axis is
in the direction of Galactic rotation, and the $z$ axis is
directed to the north Galactic pole.

In the case of an exponential density distribution, the observed
distribution of objects along the $z$ coordinate axis is described
by the expression
 \begin{equation}
  N(z)=N_1 \exp\biggl(-{|z-z_\odot|\over h_1}\biggr).
 \label{elliptic}
 \end{equation}
If the model of a self-gravitating isothermal disk is used for the
density distribution, then the observed distribution of objects
along the $z$ axis is described by
 \begin{equation}
  N(z)=N_2{\hbox { sech}}^2\biggl({z-z_\odot\over\sqrt2~h_2}\biggr).
 \label{self-grav}
 \end{equation}
Finally, the observed distribution of objects along the $z$ axis
for the Gaussian model is described by the formula
 \begin{equation}
  N(z)=N_3\exp\biggl[-{1\over 2}\biggl({z-z_\odot\over h_3}\biggr)^2\biggr].
 \label{Gauss}
 \end{equation}
In (\ref{elliptic})--(\ref{Gauss}) $N_i,$ $i=1,2,3,$ are the
normalization coefficients, $z_\odot$ is the offset of the
symmetry plane relative to the Sun, and $h_i$ are vertical disk
scale parameters (in case of $i=1$ is so called ``scale height'').

\sectionb{3}{DATA}\label{Data}

(1) We use the data on known OB associations with reliable
distance estimates from Mel'nik \& Dambis (2009). The distances
derived by these authors were reconciled with the Cepheid scale.
The catalog contains 91 OB associations; the distances to them do
not exceed 3.5~kpc.

(2) We use a sample of Galactic masers with measured trigonometric
parallaxes. These sources, associated with very young stars, are
located in regions of active star formation. Highly accurate
astrometric VLBI measurements of the trigonometric parallaxes and
proper motions have already been performed for more than 130 such
masers by several teams in the USA, Japan, Europe, and Australia
(Reid et al. 2014; Bobylev et al. 2016; Rastorguev et al. 2016).
The error in the stellar parallax determined by this method is, on
average, less than 10\%. For these objects we use the following
restriction on heliocentric distance $r<6$~kpc. For all objects we
use the restriction on $r$ in order to avoid the influence of
bending of the disk (Bobylev 2013).

(3) We supplemented the well-known catalog of HII regions (Russeil
2003) with new photometric estimates of the distances to several
star-forming regions collected in Russeil et al. (2007) and
Mois\'es et al. (2011). From this catalog we took only those
distances that were obtained photometrically. For these objects we
use the restriction $r<4.5$~kpc.

(4) We use the sample of classical Cepheids belonging to our
Galaxy described by Mel'nik et al. (2015). This sample contains
674 stars whose distances were determined from the most recent
calibrations using both optical and infrared photometric
observational data. We have two samples of Cepheids with mean ages
$\approx75$ and $\approx138$~Myr. For these objects we use the
restriction $r<4$~kpc.

According to Bobylev \& Bajkova (2014), we took a
$6.2\times1.1$~kpc rectangle oriented at an angle of $-13^\circ$
to the $y$ axis and displaced from the Sun by 0.3~kpc toward the
Galactic anticenter as the simplest Local-arm model (see Fig.~1--2
in Bobylev \& Bajkova 2016b). For a sample of masers with
measured trigonometric parallaxes, we use the division into spiral
arms according to Reid et al. (2014).

 \begin{table}[t]                                     
 \caption[]{\small
The offset of the symmetry plane relative to the Sun $z_\odot$ and
the vertical disk scale height $h_{i},$ $i=1,2,3$ obtained using
models~(1), (2) and (3) from samples with trigonometric,
spectrophotometric or ``period-luminosity'' distances
  }
  \begin{center}  \label{t:1}
  \small
  \begin{tabular}{|c|c|c|c|l|c|}\hline
 $z_\odot,$~pc & $h_1,$~pc & $h_2,$~pc & $h_3,$~pc & Sample   \\\hline
 $-18\pm5$  &   $51\pm5$   &   $46\pm5$   &   $54\pm5$   &  69 masers with parallaxes \\
 $-13\pm4$  &   $46\pm3$   &   $36\pm3$   &   $45\pm3$   &  59 OB associations \\
 $-14\pm4$  & $46.6\pm2.8$ & $35.6\pm2.7$ & $48.4\pm2.8$ & 147 HII regions \\
 $-26\pm2$  & $64.1\pm2.4$ & $52.1\pm1.9$ & $68.2\pm2.5$ & 195 young Cepheids \\
 $-26\pm2$  & $83.1\pm2.4$ & $72.0\pm2.3$ & $85.2\pm2.8$ & 192 old Cepheids \\\hline
 \end{tabular}\end{center}
 \end{table}

\sectionb{4}{RESULTS AND DISCUSSION}\label{Results}
In Table~1 the offset of the symmetry plane relative to the Sun
$z_\odot$ and the vertical disk scale height $h_{i},$ $i=1,2,3$
obtained using models~(1)--(3) are given. These parameters and
their errors were found by fitting the models to the histograms
and through Monte-Carlo simulations. For this purpose, we
constructed the histograms with a step of 20 pc in $z$ coordinate.
As can be seen from the Table~1 the values of $h_1$ and $h_3$ are
close each other, so in the Figures we quote only two lines: for
models~(1) and (2).

The histogram in the left panel of Fig.~1, shown  without fill,
was built using the whole sample of 113 masers ($r<6$~kpc). The
histogram in the left panel of Fig.~1, shown with fill, was
constructed using 44 masers of the Local arm. As clearly seen from
this figure, the maximum value of the distribution of masers from
the Local arm falls on the positive $z$. Moreover, it is
noticeable the strong influence of the Local arm masers on the
distribution of the whole sample, so we make a conclusion about
the need to exclude objects of the Local arm. On the right panel
of Fig.~1 it is presented a histogram constructed using 69 masers,
without objects of the Local arm.

Histograms built on OB associations, HII regions, young and old
Cepheids are given in Fig.~2. From the comparison with the results
obtained on these samples, without exception Local arm objects
(Bobylev \& Bajkova 2016a), can see a significant improvement in
the histograms (they become closer to the Gaussian) of masers and
OB associations. Previously (Bobylev and Bajkova 2016a), we have
found the following vertical disk scale heights:
 $h_2=40.2\pm2.1$~pc from 91 OB associations,
 $h_2=48.4\pm2.5$~pc from 187 HII regions,
 $h_2=60.1\pm1.9$~pc from 246 young Cepheids, and
 $h_2=72.5\pm2.3$~pc from 250 old Cepheids.
We see that after the exception objects Local arm, the value of
$h$ is always decreasing.

\begin{figure}[!tH]
 \label{f1}
 \vbox{ \centerline{\psfig{figure=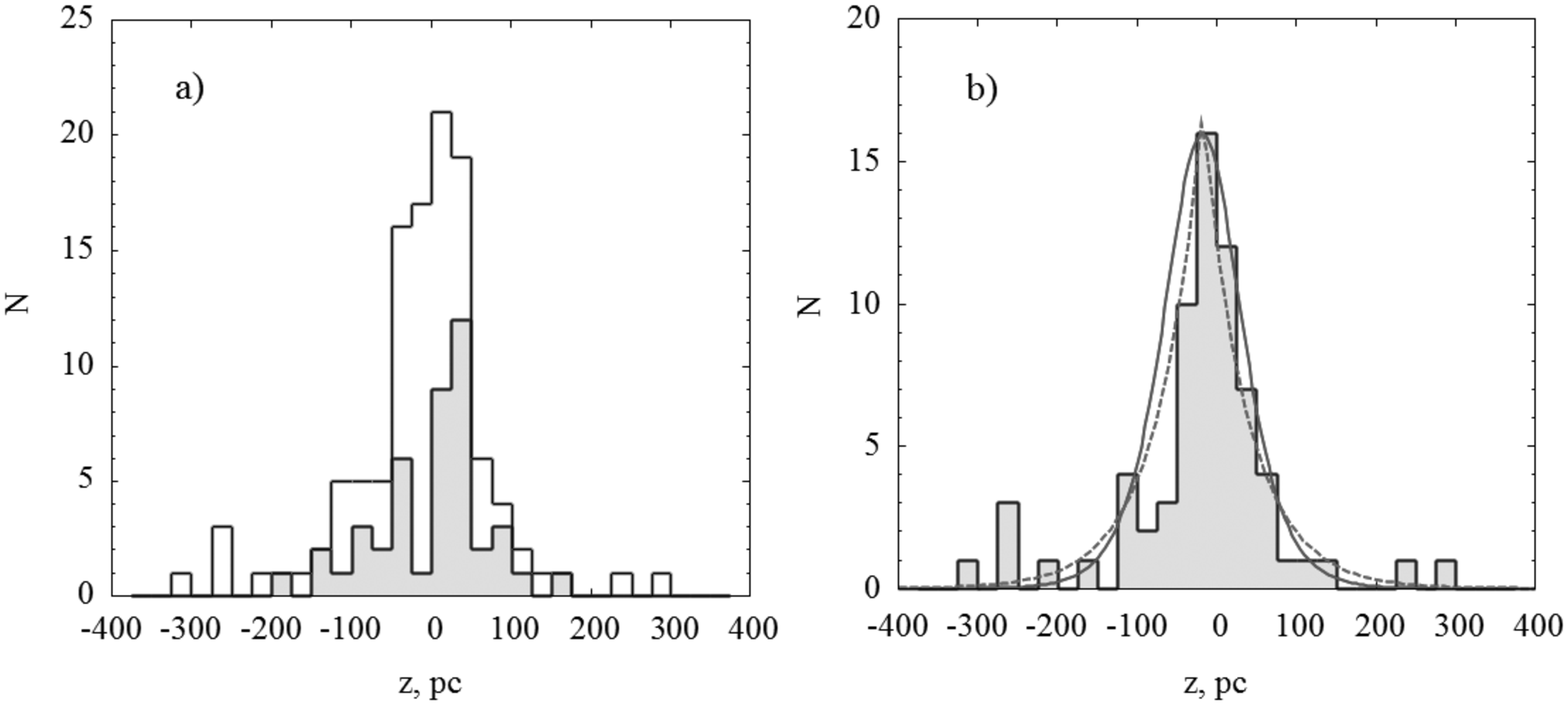,width=118mm,angle=0,clip=}}
 \vspace{1mm} \captionb{1}
{Histograms of the z distribution of masers:  the whole sample
(without fill pattern), masers belonging to the Local arm (fill
pattern) (a), the sample without masers belonging to the Local
arm, the dashed and solid lines represent models~(1) and (2),
respectively (b).
 }}
 \end{figure}
\begin{figure}[!tH]
 \vbox{ \centerline{\psfig{figure=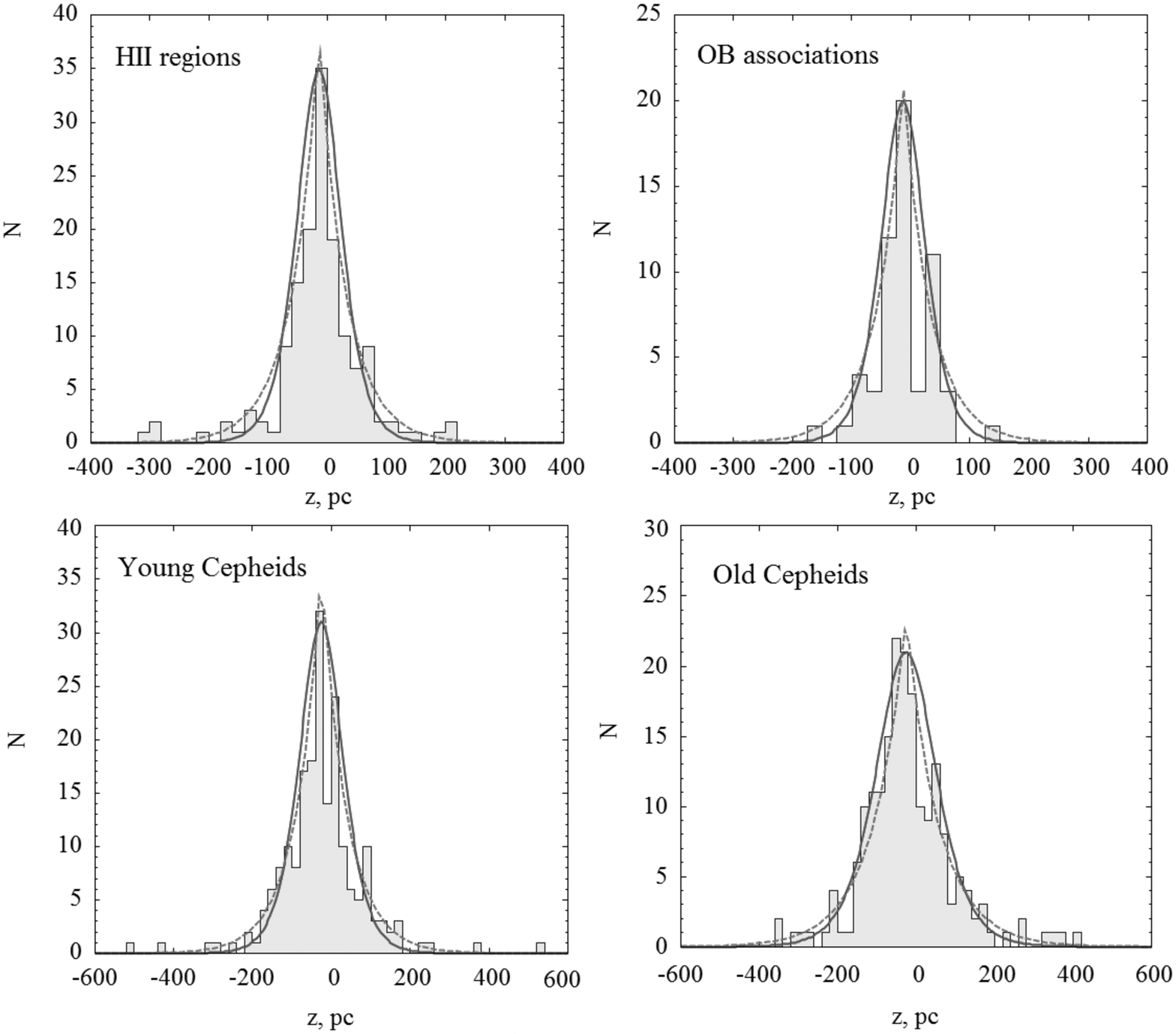,width=118mm,angle=0,clip=}}
  \label{f2}
 \vspace{1mm} \captionb{2}
{Histograms of the $z$ distributions of 59 OB associations, 147
HII regions, 195 young Cepheids and 192 old Cepheids. The dashed
and solid lines on all panels represent models~(1) and (2),
respectively.
 }}
\end{figure}
\begin{figure}[!tH]
 \label{f3}
 \vbox{ \centerline{\psfig{figure=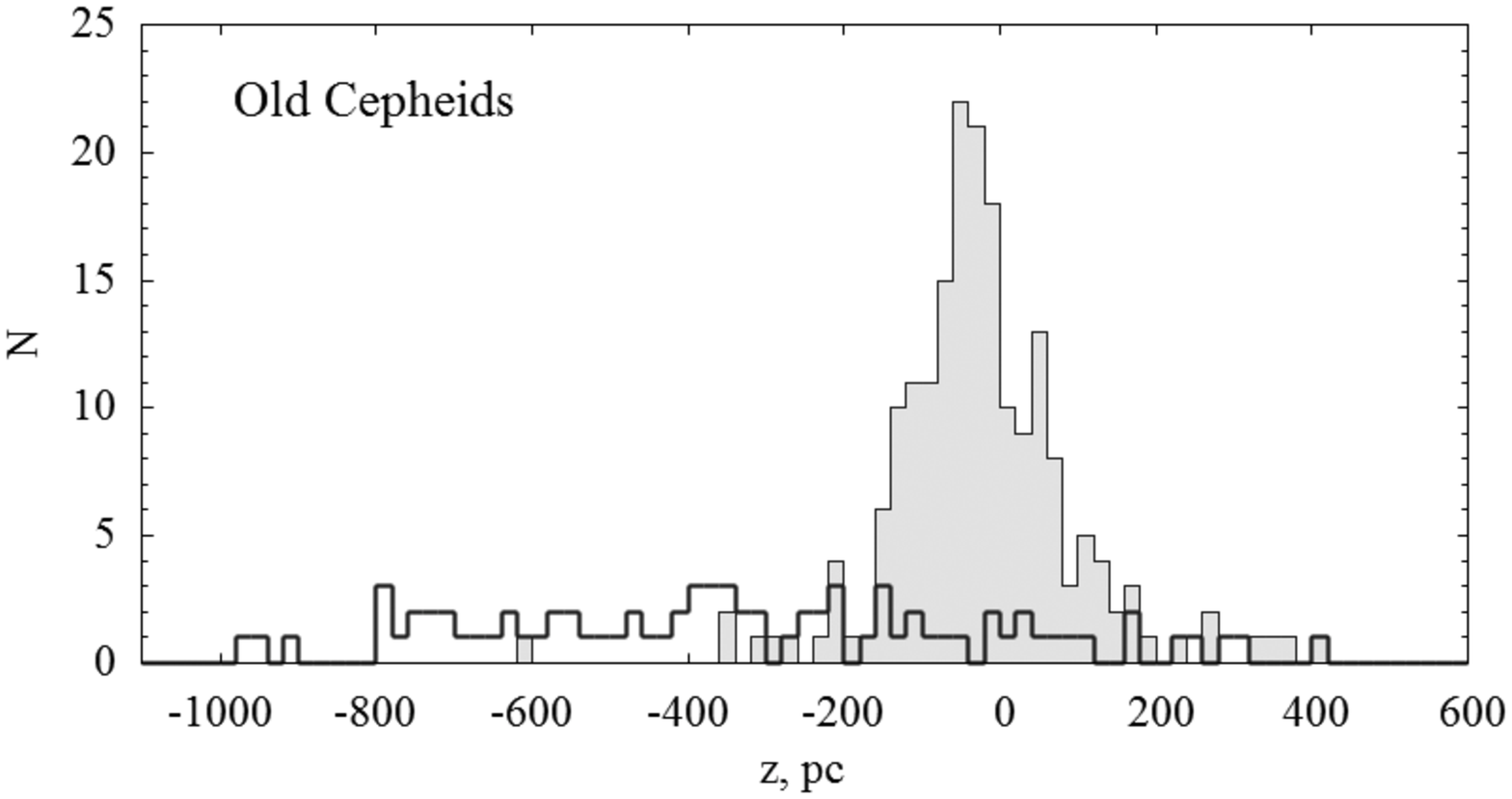,width=98mm,angle=0,clip=}}
 \vspace{1mm} \captionb{3}
{Histograms of the z distribution of old Cepheids: the sample with
heliocentric distance $4<r<10$~kpc (without fill pattern) and the
sample with distances $r<4$~kpc (fill pattern).
 }}
 \end{figure}

Bobylev \& Bajkova (2016b) have obtained the following estimates
from objects located in the inner region of the Galaxy: $z_\odot=
-5.7\pm0.5$~pc and $h_2=24.1\pm0.9$~pc from the sample of 639
methanol masers, $z_\odot=-7.6\pm0.4$~pc and $h_2=28.6\pm0.5$~pc
from 878 HII regions, $z_\odot=-10.1\pm0.5$~pc and
$h_2=28.2\pm0.6$~pc from 538 giant molecular clouds.

A detailed overview of the parameters $z_\odot$ and $h$, obtained
by different authors, can be found in work of Bobylev \& Bajkova
(2016a). Here we only note the result of the analysis of the
extensive sample of giant molecular clouds (the ATLASGAL program)
with kinematic distances from the work of Wienen et al. (2015).
These authors found $z_\odot\sim-7\pm1$~pc and $h\sim28\pm2$~pc.

It is interesting to note that the broad wings on the histograms
of z distributions are observed practically in each of our samples
(see Fig.~1--2), with the exception of OB associations. The wings
are also visible in the histogram of Wolf-Rayet stars (Fig.~2 in
Bobylev \& Bajkova 2016a) and methanol masers (Fig.~6 in Bobylev
\& Bajkova 2016b). Thus we see that (i)~the histogram wings are
always better approximated by model~(1), (ii)~all three models do
not end up describing the wide wings.

A detailed study of all effects leading to the formation of wide
wings on the histograms is the subject of a separate research due
to the great complexity of the problem. Here we would like to
mention only one of the possible causes associated with the warp
of the galactic disk (Bobylev 2013). For this, we use the old
Cepheids. In Fig.~3 there are two histograms built using two
different samples. The first sample consists of 192 Cepheids,
located at distances $r<4$~kpc (also shown in Fig.~2) and the
second sample includes 81 Cepheids of the distance interval
$4<r<10$~kpc. As can be seen from Fig.~3, the distribution of
distant Cepheids (second sample) is considerably shifted toward
the negative $z$ and has a much bigger variance compared to the
distribution of the Cepheids of the first sample: it can be
described, for example, on the basis of model~(3) with the
following parameters: $z_\odot=-321$~pc and $h_3=241$~pc. Thus, we
have two distributions with very different properties. So the
presence of broad wings on the histograms of z-distributions is
possible due to the penetration of objects with properties of the
second sample in the first sample due to large measurement errors
of the distances of the objects.

 \sectionb{5}{CONCLUSIONS}\label{conclusions}
To study the vertical distribution of visible matter in the thin
disk of the Galaxy we used several samples of young objects from
the solar neighbourhood of radius no more than 4.5~kpc (for masers
no more than 6~kpc). In the sample of masers with measured
parallaxes, it is shown that in this task, it is better not to use
the objects of the Local arm.

To each of these samples, we applied three models of the density
distribution: the model of an exponential distribution, the model
of a self-gravitating isothermal disk, and the model of a Gaussian
density distribution. The results obtained based on models~(1) and
(3) are very close.

Based on the model of an exponential distribution, we have found
the following vertical disk scale heights:
 $h_1=51\pm5$~pc from 69 masers with trigonometric parallaxes,
 $h_1=46\pm3$~pc from 59 OB associations,
 $h_1=46.6\pm2.8$~pc from 147 HII regions,
 $h_1=64.1\pm2.4$~pc from 195 young Cepheids (Age$\approx$75~Myr), and
 $h_1=83.1\pm2.4$~pc from 192 old Cepheids (Age$\approx$138~Myr).

Based on the model of a self-gravitating isothermal disk, we have
found the following vertical disk scale heights:
 $h_2=46\pm5$~pc from masers,
 $h_2=36\pm3$~pc from OB associations,
 $h_2=35.6\pm2.7$~pc from HII regions,
 $h_2=52.1\pm1.9$~pc from young Cepheids, and
 $h_2=72.0\pm2.3$~pc from old Cepheids.

 \thanks{
We are grateful to the referee for his helpful remarks that
contributed to an improvement of this paper. This study was
supported by the ``Transient and explosive processes in
astrophysics'' Program of the Presidium of Russian Academy of
Sciences (P--7).
 }

 \References
  \refb Bobylev V. V., 2013, Astron. Lett., 39, 753 
  \refb Bobylev V. V., Bajkova A. T.,  2014, Astron. Lett., 40, 783
  \refb Bobylev V. V., Bajkova A. T., 2016a, Astron. Lett., 42, 1
  \refb Bobylev V. V., Bajkova A. T., 2016b, Astron. Lett., 42, 182
  \refb Bobylev V. V., Bajkova A. T., and Shirokova K. S., 2016, Balt. Astron., 25, 15
  \refb Mel'nik A. M., Dambis A. K., 2009, MNRAS, 400, 518
  \refb Mel'nik A. M., Rautiainen P., Berdnikov L. N. et al., 2015, AN, 336, 70
  \refb Mois\'es A. P., Damineli A., Figueredo E. et al., 2011, MNRAS, 411, 705
  \refb Rastorguev A. S., Zabolotskikh M. V., Dambis A. K. et al., 2016, arXiv: 1603.09124
  \refb Reid M. J., Menten K. M., A. Brunthaler A. et al., 2014, ApJ, 783, 130
  \refb Russeil D., 2003, A\&A, 397, 133
  \refb Russeil D., Adami C., and Georgelin Y. N., 2007, A\&A, 470, 161
  \refb Wienen M., Wyrowski F., Menten K. M. et al., 2015, A\&A, 579, 91

  \end{document}